\documentclass{article}
\usepackage{graphicx}
\usepackage{amsmath}

\input{tcilatex}

\begin{document}

\begin{center}
{\LARGE Superluminal Phenomena as Instantaneous Transferring of Excitations
in Correspondence with the Wigner Principle of Causality}

\textbf{Mark E. Perel'man }$^{\ast)}${\large \ }

(\textit{Racah Institute of Physics, Hebrew University, Jerusalem, Israel) }
\end{center}

\bigskip

It is shown that the transferring of light signal faster-than-c can take
place exclusively as the instantaneous quantum tunneling. Its extent (space
size of instanton) must be inversely relative to lack of energy till nearest
stable or resonant state. ''Superluminality'' leads to nonlocality ''in
small'', in a near field zone, and can be described by 4-vector $A_{\mu}$,
the $\mathbf{E}$ and $\mathbf{H}$ fields remain local.

These assertions are proven, independently, in the frameworks of general
field theory and by the theory of time duration of scattering, with taking
into account the entropy growing. They are correlated with peculiarities of
tunneling and frustrated total reflection. The results correspond to the
Wigner formulation of causality: ''The scattered wave cannot escape a
scatterer before the initial wave reaches it'' and are conformed to
experimental data on superluminal transferring.


\qquad

{\Large 1. INTRODUCTION}

In the experiments [1-10] has been observed that the light signal could be
transferred on definite distances via specially chosen media with speed
greater vacuum light velocity (''superluminal'' or ''light faster-than-c'').
Almost in all theoretical researches it has been adopted, from the outset,
that such exotic speed is unreal, primary impossible, and therefore these
observations would be explainable by some reformulation of interference
conditions, by account of reshaping of pulses form and so on (e.g. [11-13]).

However, these experimental data can be considered as some manifestation of
reality of the superluminal phenomena. Such point of view requires search of
the exclusive conditions and opportunities at which superluminal phenomena
are not absolutely impossible. Therewith such possibilities must be revealed
on the most general basis without introduction of any additional hypotheses,
and just such approach is the main purpose of the paper.

Let us begin with overview of the experimental data. Superluminal
transmission of light pulses (or of its forward front only?) through some
media has been registered in rarefied gas [1] (the most impressive
observations of this effect have been made nearby the resonant frequencies
[2, 3]), as well as under propagation of microwaves through the air (several
experiments of two independent groups [4]). The superluminal speed of light
pulses has been registered also at their propagation through solid-state
films [5] and interfering filters [6], at frustrated total internal
reflection (FTIR [7]). It results in some features of diffraction pictures
[8] and even the speed of front of ionization can be superluminal [9].
Moreover, in electronic systems such phenomena can be observed also [10].
Reviews of many experimental data and some theoretical models are given in
[11] and in the proceedings of conferences [12], but we do not consider
these models here [13].

These observations are indicative for the reality of the superluminal
propagation or of signal transferring, at least along comparatively small,
characteristic distances, and require a reasonable theoretical
interpretation. Notice that to these observations must be added the long
time noted discrepancy with results of tunneling calculations that show
negative duration of such transitions (e.g. [14]).

How the strict restriction of signal speed over the large distances can be
reconciled with these observations?

The regular measurements of signal speed were carried out in the far field,
i.e. at distances, which are by far larger than characteristic scales,
determined by the wavelengths and/or the uncertainty relations. Hence the
superluminal phenomena could be only attributed to such scales that are
negligible in the far field measurements, i.e. they should be limited by a
near field zone. The energy propagation (group) speed determined by the
Poynting's vector in far field could remain subluminal.

The question under consideration arises from the beginning for some
specification of the causality principle itself. Its conventional statement: 
\textit{''No outgoing signal can leave a point before it is reached by the
initial one''} is essentially classical. It is not applicable literally to
the quantum theory of measurements, does not admitting so exact localization
of the area of absorption and/or emission. Thereby, one has to adopt the
weaken formulation given by Wigner [15]: \textit{''The scattered wave can
not leave a scatterer before the initial wave reaches it''}, which can be
viewed as the general (or quantum) causality principle. It must be noted
that the Wigner principle is reconcilable, in particular, with occurrence of
some items distinct from zero outside of the light cone in the causal
propagator - just they are suggestive for the appearance of superluminal
phenomena [16].

The singularity and apparent exotic character of a problem requires its
research from different standpoints. Therefore we shall begin our
consideration from the rigorous mathematical, but sufficiently general
investigation of an opportunity of existence and properties of the
superluminal phenomena in the scope of general field theory, without
concretization of media, in which can be observed such phenomena (Section 2,
the Appendix A).

Then we consider an opportunity of description of superluminal phenomena
within the framework of general theory of time delay with taking into
account restrictions, which are imposed on parameters of processes by the
requirements of entropy growth (Section 3-1, the Appendix B).

The opportunities of confirmation of general results are considered within
the quantum electrodynamics (QED, Subsection 3-2), in the theory of
tunneling (Subsection 3-3) and, in particular, in the phenomenon of the
frustrated total internal reflection (FTIR, Subsection 3-4) as the most
known and the most old example of tunneling. Then in the Section 4 we carry
out comparison of our results with experimental data. Some possible models
and interpretations of results are discussed in the Section 5 and thereby
are outlined some possibilities of further investigations.

As the method used in the Section 2 is not traditional one, let's
forewarning it by such remarks. We examine passage of a signal throw linear,
homogeneous, stationary medium (without its further concretization). This
process is described by the integral relationship:

$O(t,\mathbf{r})=\int dt^{\prime}d\mathbf{r}^{\prime}R(t-t^{\prime},\mathbf{r%
}-\mathbf{r}^{\prime})\ I(t^{\prime},\mathbf{r}^{\prime})$. (1.1)

The set of response functions $\{R(t,\mathbf{r})\}$ of this system may be
divided formally on two classes: the set of local functions $\{R_{L}(t,%
\mathbf{r})\}$, for which $ct\geq r$, and the set of nonlocal functions $%
\{R_{NL}(t,\mathbf{r})\}$, which are characterized by nonequality $ct<r\ $%
and thereby described the superluminal signal (energy) transferring only.
The analysis of such rigorous division of functions of response onto two
classes shows the exact limitation of peculiarities of the class of nonlocal
functions.

The further analysis of this class of functions will showed, that the
instant transfer of signal can occur only onto the certain distances $a$,
which must be determined by the difference between the energy of scattered
particle and the energy of nearest stable or resonant state. Moreover, such
transition can take place exclusively by the tunneling, i.e. they present a
pure quantum phenomenon and corresponding response function does not depend
on magnitudes of vectors $\mathbf{E}$ and $\mathbf{H}$. Therefore, just as
in the effects of Aharonov-Bohm [17], Casimir (e.g. [18]) and so on,
superluminal phenomena should be described via potential $A_{\mu}$ only.

All subsequent items demonstrate concretization and illustrations of these
principal results. Notice that the outstripping of particles on completely
determined distances can be described as the ''gain duration'' in its
kinematics and is not reflected on properties of far field.

\qquad

{\Large 2. PERMISSIBLE FORMS OF RESPONSE FUNCTIONS}

\qquad

{\large 2-1. LOCAL AND NONLOCAL INTERACTIONS}

The method of division of response functions on various classes is based on
usage of projectors of properties of physical systems introduced by von
Neumann [19] for description of qualitative (alternative) properties of
systems, which are completely present or are completely absent, such as
causality, locality, mass-spectrality, definite symmetries, etc.

In the quantum mechanics projectors of type $P_{\psi }$ $=$ $\left| \psi
\right\rangle \left\langle \psi \right| $ are used for separation of certain
states of systems in the Hilbert space, on the base of which are constructed
matrices of density [20], etc. Unlike it, we shall consider projectors of
allowable supports of response functions (areas of change of variables,
outside of which functions equal zero) in the usual time-space, the
corresponding method was proposed and developed in [21, 22].

Projectors of locality (of light cone) $P_{L}$ and of nonlocality (the areas
outside of light cone) $P_{NL}$ are, accordingly, represented as

$P_{L}(x)=\theta(x^{2})\equiv\theta(t^{2}-r^{2}),\qquad P_{NL}(x)=1-\theta
(x^{2})\equiv\theta(-x^{2})$. (2.1)

Any function can be decomposed on local and nonlocal parts: $%
R(x)=R_{L}(x)+R_{NL}(x)$. These parts can be separated from any function by
the appropriate projector:

$R_{L}(x)=P_{L}(x)\ R(x),\qquad R_{NL}(x)=P_{NL}(x)\ R(x)$ . (2.2)

The condition of completeness, $P_{L}+P_{NL}=1$ (see the Appendix A), allows
to deduce from (2.2) the equations of orthogonality for these functions:

$P_{NL}(x)\ R_{L}(x)=0$; (2.3)

$P_{L}(x)\ R_{NL}(x)=0$, (2.4)

i.e. the projector of light cone $P_{L}$ is orthogonal to functions with
support out of the cone, etc.

From the law of energy conservation for classical systems or from the
principle of unitarity for quantum systems follows that functions, which
describe input and output signals, are square integrable. Therefore for them
and for the function of medium response $R(t,\mathbf{r})$ exist the complete
and partial Fourier-transforms (in the rigged Hilbert space [21] at least):

$O(\omega,\mathbf{k})=R(\omega,\mathbf{k})\ I(\omega,\mathbf{k})$ (1.1')

(functions and their Fourier-transforms are designated by the same symbol
and differ by arguments). The Fourier-transformation of (2.3) to $(\omega ,%
\mathbf{k})$- representation with simultaneous account of causality (the
projector $P_{C}(x)=\theta(-t)$, i.e. at allocation only the bottom part of
light cone) leads to relativistic generalization of Kramers-Kr\"{o}nig
dispersion relations of complex structure [22], though admitting the
approximated forms [23].

Let's consider these equations of restriction in other variables. In
variables $(t,\mathbf{k})$ the projector gets a kind:

$P_{L}(t|\mathbf{k})\equiv\ $\textbf{F}$_{\mathbf{k}}[P_{L}(t,\mathbf{r})]=(%
\frac{1}{2\pi})^{3}\int d\mathbf{r}e^{-i\mathbf{kr}}\theta(t^{2}-r^{2})=%
\frac{1}{2\pi^{2}k^{3}}[\sin(kt)-kt\cos(kt)]$. (2.5)

Hence, $P_{L}(t|\mathbf{k})\rightarrow\delta(\mathbf{k})$ at $t$ $%
\rightarrow\infty$, and at $t$ $\rightarrow0$ it is presented as

$P_{L}(t|\mathbf{k})\rightarrow\frac{1}{2\pi^{2}k^{3}}\dsum _{1}^{\infty}$ $%
(-1)^{n+1}\ n\ (kt)^{2n+1}$. (2.6)

The Fourier transformation of (2.3) leads to the integral equation for local
functions in these variables:

$R_{L}(t|\mathbf{k})$ $=\int d\mathbf{q\ }P_{L}(t|\mathbf{q})$ $R_{L}(t|%
\mathbf{k-q})$, (2.7)

from which, via (2.6) and so on, some asymptotic estimations can be received.

The research of features of completely nonlocal transfer functions on the
basis of (2.4) is the most interesting. The Fourier transform of this
equation results in the equation:

$\int d\mathbf{q\ }P_{L}(t|\mathbf{q})$ $R_{NL}(t|\mathbf{k-q})=0$. (2.8)

As its feasibility does not depend on magnitudes of $t$ and $\mathbf{k}$, it
is reduced to the algebraic equation:

$P_{L}(t|\mathbf{q})$ $R_{NL}(t|\mathbf{k-q})=0$, (2.9)

and its equality to zero is determined by features of $P_{L}(t|\mathbf{q})$
as function of variable $t$.

On the basis of fundamental definition $\xi \times \delta (\xi )=0$ can be
shown that if function $P(x)$ is nonsingular with only isolated zeros in the
points $x_{m}$ (not higher $n$-th order), the general solution of the
algebraic equation $P(x)\times f(x)=0$ is of the such form:

$f(x)=A\ \delta(P(x))=\sum_{0}^{n}\sum_{m}a_{n,m}\ \delta^{(n)}(x-x_{m})$
(2.10)

with arbitrary factors $a_{n,m}$, which other reasons or models may
determine.

Hence, the solution of (2.9) has the form:

$R_{NL}(t|\mathbf{k,q})=f(t|\mathbf{k,q})\ \delta(P_{L}(t|\mathbf{k}))$,
(2.11)

and it is necessary to determine in the argument of $\delta $-function its
zeros as function of temporary variable. As in the expression (2.5) the
values $kt=2\pi n$ are no its roots, from the argument of $\delta $-function
can be factored out $\cos (kt)/2\pi ^{2}k^{3}$,$\ $and thus in (2.11)
remains $\delta (\tan (kt)-kt)$. The first numerical roots of the
transcendent equation $\tan x=x$ are equal to $0;4.493;7.725;$
(approximately). As it will be shown below, the minimal extent of path at
nonlocal interaction is equal to the wavelength $\lambda $, therefore if $%
kt=b$, the maximal speed of pulse transfer must be $u\approx \lambda /t=2\pi
c/b$. Hence, at decomposition of the argument of $\delta $-function all
roots, except the first two, can be omitted.

In accordance with (2.6) the zero in point $t=0$ is of the third order and
consequently the general solution of (2.11) must be written down as

$R_{NL}(t|\mathbf{k,q})=A\delta(P_{L}(t|\mathbf{k,q}))$ $=\varphi_{0}%
\delta(t)+\varphi_{1}\delta^{\prime}(t)+\varphi_{2}\delta^{\prime\prime
}(t)+\varphi_{3}\delta(t-4.5/k)$ $+...$ .(2.12)

with arbitrary functions $\varphi_{n}$ $=\varphi_{n}(\mathbf{k,q})$. Last
term of (2.12) results to $\lambda/t\ \symbol{126}\ 1.4c$ and apparently is
of type of harbinger or aftershock and therefore further is not examined.

Other terms of (2.12) show that if there is the nonlocal transfer of
interaction, it should occur instantaneously, without dependence on time and
consequently must be described by the stationary Laplace equation. The
substitution of $R_{NL}(t,\mathbf{r})$ in (1.1) leads to the expression

$O_{NL}(t,\mathbf{r})=\dint d\mathbf{r}^{\prime}[f_{0}(\mathbf{r}-\mathbf{r}%
^{\prime})+f_{1}(\mathbf{r}-\mathbf{r}^{\prime})(\partial/\partial t)+f_{2}(%
\mathbf{r}-\mathbf{r}^{\prime})(\partial/\partial t)^{2}]\ I(t,\mathbf{r}%
^{\prime})$, (2.13)

where functions $f_{n}(\mathbf{r)}$, which describe medium, are not
concretized. The expression (2.13) show that the nonlocal causal response of
system depends on spatial properties of medium and on velocity and
acceleration of the source changing, i.e. on forces operating in it.

The instantaneous transferring of all three characteristics of process is
necessary and is sufficient for complete restoration of the initial form and
dynamics of process. Three spatial transfer functions included in (2.13) can
be considered as independents, concerning, accordingly, to static,
kinematical and dynamic parts of nonlocal field.

\qquad

{\large 2-2. DISTANCES OF INSTANTANEOUS JUMPS OF PHOTONS}

Let's consider the projector describing instant jump of particle into the $%
x\ $direction on distances not smaller $a$. Projectors of such property
should include the bottom part and the top part of light cone remote at the
moment $t=0$ on distance $a$:

$P_{a}(t,x)=\theta(-t)\theta(t^{2}-x^{2})+\theta(t)\theta(t^{2}-(x+a)^{2})$.
(2.14)

This projector separates within the whole set of response functions those
functions, which correspond to jump on distances not smaller $a$:

$R_{a}(t,\mathbf{r})=P_{a}(t,x)\ R(t,\mathbf{r})$. (2.15)

Thereby they should satisfy the equation of orthogonality:

$(1-P_{a}(t,x))\ R_{a}(t,\mathbf{r})=0$. (2.15')

Fourier-transform of the projector in (2.15') as function of temporary
variable has the form:

F$_{\omega}[1-P_{a}(t,x)]=\frac{1}{2\pi i\omega}(e^{-i\omega|x|}-e^{-i%
\omega|x+a|})$, (2.16)

and the response function of system describing instant jumps on distances $%
\QTR{sl}{l}\ \geq a$ is of type

$R_{a}(\omega|\mathbf{r})=f(\omega|\mathbf{r})\ \omega\ \delta(ie^{-i\omega
|x+a|}-ie^{-i\omega|x|})$. (2.17)

The zeros of argument of $\delta$-function correspond to the equality $%
\omega a=2\pi n$. Thus the spatial extent of instant jumping can be equal to
one or several wavelengths:

$a=n\lambda,\qquad n=1,2,3,...,$ (2.18)

i.e. exceeds, already with $n=1$, the meaning of uncertainty ($\Delta x\
\Delta k\geq%
%TCIMACRO{\UNICODE{0xbd}}%
%BeginExpansion
{\frac12}%
%EndExpansion
$, $\Delta x\approx a$, $\Delta k\approx2\pi/\lambda$). Thereby the ''gain
time'' at the jump is equal to $\Delta T=n\ \lambda/c$ and can be
measurable. (Such possibility was announced in [24].)

If the considered process is determined by the difference between energies
of scattered particle and of stable (or resonant) state $\Delta\omega
=\omega-\omega_{0}$, it is necessary to carry out, already in (2.15), the
appropriate subtraction:

$R_{a}(\omega|\mathbf{r})-R_{a}(\omega_{0}|\mathbf{r})\rightarrow
R_{a}(\Delta\omega|\mathbf{r})$. (2.19)

The replacement $\omega\rightarrow\Delta\omega$ in (2.16) leads to the
length of instant tunnel jump:

$a=(2\pi c/\Delta\omega)\ n,\qquad n=1,2,3,...$, (2.18')

that corresponds, at $n=1$, to the distance or to the ''gain time'',
measured in the experiments [2].

\qquad

{\large 2-3. ENERGY-MOMENT RELATION }

The interaction functions (matrix elements of transition) in variables $%
(\omega,\mathbf{k})$ can be naturally divided, up to an exit on the mass
surface, onto two classes: $R_{E}(\omega,\mathbf{k})$, at which $|\omega
|>\left| \mathbf{k}\right| $, i.e. there is a surplus of energy relative to
moment, and $R_{NE}(\omega,\mathbf{k})$, at which $|\omega|<\left| \mathbf{k}%
\right| $, i.e. the energy is lesser than value appropriate to moment.

By introducing the projectors of 4-cone $(\omega,\mathbf{k})$,

$P_{E}(\omega ,\mathbf{k})=\theta (\omega ^{2}-k^{2})$, $\qquad
P_{NE}(\omega ,\mathbf{k})=\theta (k^{2}-\omega ^{2})\equiv 1-P_{E}(\omega ,%
\mathbf{k})$, (2.20)

we receive, by the exact analogy with (2.3-4), the equations of
orthogonality:

$P_{NE}(\omega,\mathbf{k})\ R_{E}(\omega,\mathbf{k})=0$, (2.21)

$P_{E}(\omega,\mathbf{k})\ R_{NE}(\omega,\mathbf{k})=0$. (2.22)

The Fourier transforms of these projectors to variables $(t,\mathbf{k})$ are

$P_{NE}(t|\mathbf{k})=\frac{1}{2\pi}\int_{-k}^{k}d\omega\ e^{-i\omega t}=%
\frac{\sin(kt)}{\pi t}$, (2.23)

$P_{E}(t|\mathbf{k})=\delta(t)-P_{NE}(t|\mathbf{k})$. (2.23')

The Equation (2.21) results, in view of the condition of causality ($f_{E}(t|%
\mathbf{k})=0$ at $t>0$), in the integral relation:

$\dint _{-\infty}^{t}dt\prime\ R_{E}(t-t\prime|\mathbf{k})\frac{%
\sin(kt\prime)}{\pi t\prime}=0$. (2.24)

As integrand does not depend on magnitudes of $t\ $and $k$, it should be
equal to zero, and its solution, as well as above, can be expressed through $%
\delta $-function depending on $t$ and $k$:

$R_{E}(t|\mathbf{k})=f(t|\mathbf{k})\ \delta(\sin(kt)/t)=f(t|\mathbf{k})\ t\
\delta(\sin(kt))$. (2.25)

Delta-function of periodic argument is represented by series: $\delta(\sin
x)=\sum_{0}^{\infty}\delta(x-\pi n)$, and consequently the solution of
(2.21) can be written as

$R_{E}(t|\mathbf{k})=\pi\ t^{2}f(t|\mathbf{k})\sum_{1}^{\infty}\delta(k-\pi
n/t)$, (2.26)

in which, by virtue of the extra factor $t$ in (2.25), there is no term with 
$n=0$.

From here follows that if the energy of system exceeds the value appropriate
to moment, the instant transfer of interaction is impossible, since it
requires the infinite large moment!

The expression (2.26) corresponds, as though, to natural condition of
resonance transfer of the ideal retarded signal in any system with
superfluous energy. In the classical theory, with fixed distance between
emitter and receiver, it shows that the interaction executes by integer
numbers of half-waves.

Let's consider $(\omega,\mathbf{r})$-representation of the Equation (2.22).
As it is identical to (2.9) with replacements $(t,\mathbf{k})\rightarrow
(\omega,\mathbf{r})\ $only, then by analogy with (2.12) it must be written
that

$R_{NE}(\omega|\mathbf{r})=\varphi_{0}(\mathbf{r})\delta(\omega)+\varphi
_{1}(\mathbf{r})\delta^{\prime}(\omega)+\varphi_{2}(\mathbf{r})\delta
^{\prime\prime}(\omega)+\varphi_{3}(\mathbf{r})\delta(\omega-4.5/r)+...$
(2.27)

or, in the temporary representation,

$R_{NE}(t|\mathbf{r})=$ $f_{0}(\mathbf{r})+t\ f_{1}(\mathbf{r})+t^{2}f_{2}(%
\mathbf{r})+...$ . (2.28)

Hence, at the lacking of energy concerning value of moment (it is just the
case of tunneling) the opportunity of instant transition at $t=0\ $is not
excluded and may be described by the function$\ f_{0}(\mathbf{r}).$

Here should be specially underlined, that this nonlocality does not lead to
nonlocality of $\mathbf{E}$ and $\mathbf{H}$ fields. Such conclusion follows
from consideration of the commutators of fields, as response functions must
be expressed through Green functions:

$[E_{i}(x),E_{j}(y)]=[H_{i}(x),H_{j}(y)]=\frac{1}{4\pi i}\{\partial
_{i}\partial_{j}-\delta_{ij}\ \partial_{t}^{2}\}D(x-y)$;

$[Ei(x),Hj(y)]=\frac{1}{4\pi i}\partial_{t}\partial_{j}D(x-y)$, (2.29)

and differentiation of response function removes the unique nonlocal term $%
f_{0}(\mathbf{r})$ in (2.28).

Thus, nonlocality should be observable basically in such phenomena as the
Aharonov-Bohm effect, the Casimir effect, effects of near field optics,
etc., caused by the peculiarities of field $A_{\mu}$. Moreover, the
possibility of movement with superluminal speed is the pure quantum
phenomenon, and therefore, in the complete consent with the theory of
relativity, cannot be described in the scope of classical theory.

\qquad

{\Large 3. DURATION OF INTERACTION}

{\large 3-1. DURATION OF STATE FORMATION AND PRINCIPLE OF PASSIVITY}

The Wigner's causality admitted a negative ''durations of delay'' $\tau>-a$
and it was \textit{assumed} that $a$ can be the Compton wavelength. It seems
natural to suggest that for photons such estimation must be conformed to the
uncertainty principle and/or can be associated with an extent of near field.

Let us consider this problem in detail. The complex function of duration of
interaction can be determined by expansion of $R(\omega)$ or its logarithm
in the vicinity of some characteristic frequency of medium $\omega_{0}$ as:

$\ln R(\omega,\mathbf{r})=\ln R(\omega_{0},\mathbf{r})+i(\omega-\omega
_{0})\ \tau(\omega_{0},\mathbf{r})+%
%TCIMACRO{\UNICODE{0xbd}}%
%BeginExpansion
{\frac12}%
%EndExpansion
(\omega-\omega_{0})^{2}\ \tau^{\prime}(\omega_{0},\mathbf{r})+...\ $(3.1)

with

$\tau(\omega,\mathbf{r})\equiv\tau_{1}+i\tau_{2}$ $=(\partial/i\partial
\omega)\ln R(\omega,\mathbf{r})$. (3.2)

In this representation $\tau_{1}(\omega)$ is the time-delay during elastic
scattering (e.g. [25]) and $\tau_{2}(\omega)$ is the duration of final state
formation, some of their peculiarities are outlined in the Appendix.

The physical significance of $\tau _{2}$ becomes more transparent in its
formulation in the $S$-matrix theory as $\tau _{2}=\partial \ln |S|/\partial
\omega $. Hence, $\tau _{2}$ can be considered as a measure of temporary
non-completion of the final (free photon) state: by definition, $S$-matrix
should be unitary while describing transition between real physical states
(cf. [26]).

If we shall restrict our consideration to first two terms of (3.1), then
since for passive linear media $|R(\omega)|^{2}=|R(\omega_{0})|^{2}\ \exp
[-2(\omega-\omega_{0})\tau_{2}]\leq1$, the inequality

$(\omega-\omega_{0})\tau_{2}(\omega_{0},\mathbf{r})\geq0$ (3.3)

should be valid, and, therefore, the formation time $\tau_{2}<\ 0$ under
some values of $\omega<\omega_{0}$.

Thus, the advanced or instantaneous, in the sense of expression (1.1),
phenomena could be observed within a frequency range below the resonance
one, just in the correspondence with the Subsection 2-3.

In macroscopic theories the role of medium response function plays the
Fourier-transform of dielectric susceptibility, $\varepsilon(\omega ,\mathbf{%
r})$. Possibility of the negative sign of formation time in these theories
is emerged from such general precept: in passive media the principle of
entropy growing results in the strict inequality: $\partial(\omega
\varepsilon(\omega))/\partial\omega\geq0$ [27 ]. At substitution $%
R\rightarrow\varepsilon(\omega)-\varepsilon(\infty)=\varepsilon
_{1}+i\varepsilon_{2}$ in (3.1) this general inequality can be rewritten as

$\tau_{2}\leq\frac{1}{\omega}-\tau_{1}\frac{\varepsilon_{2}}{\varepsilon_{1}}
$. (3.4)

Hence in the region of abnormal dispersion, where must be expected
discordance between maxima of $\tau_{1}$ and $\tau_{2}$, the value of $%
\tau_{2}$ can be negative.

Temporal functions of the simplest form can be obtained in a model for
dielectric susceptibility of media with single isolated resonance:

$\varepsilon(\omega;\mathbf{r}_{1},\mathbf{r}_{2})=\rho(\mathbf{r}_{1},%
\mathbf{r}_{2})/[\omega_{0}^{2}-(\omega+\frac{i}{2}\Gamma)^{2}]$. (3.5)

After substitution in the definition (3.2), this gives at $|\omega |\ 
\symbol{126}\ \omega_{0}>>\Gamma$:

$\tau_{1}=\frac{\Gamma/2}{\delta\omega^{2}\ +\ (\Gamma/2)^{2}}+\frac{\Gamma
/2}{\varpi^{2}\ +\ (\Gamma/2)^{2}}\approx\frac{\Gamma/2}{\delta\omega ^{2}\
+\ (\Gamma/2)^{2}}$; (3.6)

$\tau_{2}$ $=\frac{2\varpi}{4\varpi^{2}\ +\ (\Gamma/2)^{2}}-\frac{\delta
\omega}{\delta\omega^{2}\ +\ (\Gamma/2)^{2}}\approx$ $\frac{1}{2\varpi}-%
\frac{\delta\omega}{\delta\omega^{2}\ +\ (\Gamma/2)^{2}}$, (3.7)

where $\delta\omega=(\omega_{0}-\omega)$, $\varpi$ $=(\omega_{0}+\omega)/2$.

Due to its conformity with the Breit-Wigner formulae the physical sense of
expression (3.6) is obvious. The expression for $\tau_{2}$\ is close to the
uncertainty value, evidently admits negative $\tau_{2}$ and could be
interpreted as advanced emission of photons or as their instantaneous jumps
onto the distance $c\left| \tau_{2}\right| $.

Making use of these expressions, the condition (3.4) transforms in

$\tau_{2}\leq\frac{1}{\omega}-\frac{\omega\Gamma^{2}}{(\delta\omega ^{2}\ +\
(\Gamma/2)^{2})(\omega_{0}^{2}\ -\ \omega^{2}\ +\ \Gamma^{2}/4)}$. (3.8)

At the margin of resonant line, $|\delta\omega|\geq\Gamma/2$, one can accept
that

$\tau_{eff}\equiv\tau_{2}-\frac{1}{\omega}\leq\ -\frac{\Gamma^{2}}{4\
\delta\omega\ (\delta\omega^{2}\ +\ (\Gamma/2)^{2})}$. (3.9)

Thus, a negative duration of formation (the ''gain of time'' at superluminal
jump) in gaseous media is possible only and only at $\omega<\omega_{0}$, and
it completely confirms results of Subsection 2-3 above (just such
frequencies were used in the quoted experiments). Moreover, the length of
superluminal ''jump'' exceeds the atomic sizes, at $c(\tau_{2}-\frac{1}{%
\omega })\ \symbol{126}\ 10^{-8}$ , only if $|\delta\omega|\leq10^{11}$,
i.e. only within the specially prepared media of a type described in [1-3].

\qquad

{\large 3-2. QUANTUM ELECTRODYNAMICS}

The causal, Stueckelberg-Feynman, propagator $D_{c}$ includes the space-like
part, nonvanishing beyond of light cone, i.e. contains the superluminal
terms. This peculiarity is usually intuitively understood as a result of
vacuum fluctuations, descriptively interpreting through uncertainty
principles.

Really, the photon propagator of the lowest order is written as

$D_{c}(t,\mathbf{r})=\overline{D}+%
%TCIMACRO{\UNICODE{0xbd}}%
%BeginExpansion
{\frac12}%
%EndExpansion
D^{(1)}\equiv\frac{1}{4\pi}\delta(c^{2}t^{2}-r^{2})+\frac{1}{2\pi i}\frac
{1}{c^{2}t^{2}\ -\ r^{2}}$, (3.10)

where the first term (the Pauli - Jordan function) is supported in the light
cone, but the second term (the Hadamard function) partially goes out the
cone, though with a fast attenuation in the outside region. Classical
electrodynamics is based on the retarded interaction or on the propagator $%
\overline{D}$ only [28] and does not include superluminal terms.

In the general case, the first term of the causal propagator is responsible
for the restriction of support of space-time parameters into the light cone, 
$c^{2}t^{2}-r^{2}\geq0$, and the second one is responsible for the
restriction of support of energy-moments parameters: $E^{2}-p^{2}c^{2}\geq
m^{2}c^{4}$. These requirements were considered in the Section 2 from
another point of view. Fulfillment of both requirements by any single
function is impossible; these two functions have different physical sense,
which can conform to different space-time effects [22]. Therefore the
durations, resulting from these two parts, may have different physical sense
and different magnitudes.

In the $(\omega,\mathbf{k})$ representation at $\eta\rightarrow+0$ the
causal propagator of the lowest order has in the Feynman gauge such form:

$D_{c}(\omega,\mathbf{k})=\frac{4\pi}{\omega^{2}\ -\ \mathbf{k}^{2}+\ i\eta}%
= $ $\frac{4\pi}{\left| \omega^{2}\ -\ \mathbf{k}^{2}\right| }\exp$ $\left( 
\frac{-i\eta}{\omega^{2}\ -\ \mathbf{k}^{2}}\right) $. (3.11)

Thereby the delay time and duration of formation are:

$\tau_{1}(\omega,\mathbf{k})=-2\pi\delta(\omega^{2}-\mathbf{k}^{2})$, (3.12)

$\tau_{2}(\omega,\mathbf{k})$ $=\frac{2\omega}{\omega^{2}\ -\ \mathbf{k}^{2}}%
\ \symbol{126}\ \frac{1}{\omega\ -\ |\mathbf{k}|}$. (3.12')

The function $\tau_{1}$ obviously describes the reemission of dressed
photon. The function $\tau_{2}$ shows that an outstripping is possible only
at the condition of certain mismatch between photon frequency and moment.
Just such mismatch takes place under conditions of the abnormal dispersion
and frustrated total internal reflection, when the absolute value of moment
grows up.

In our case, the response function (transfer amplitude) consists of the
photon propagators and the vertex parts. It can be assumed that in the
lowest order the delay and duration times at each elementary step of photon
transfer will be determined by the propagators only.

Let's continue our examinations in the $(\omega,\mathbf{r})$- representation:

$D_{c}(\omega,\mathbf{r})=-\frac{1}{4\pi r}e^{i|\omega|r}$;\qquad \ $%
D^{(1)}(\omega,\mathbf{r})=-\frac{1}{2\pi ir}\sin(\omega r)$, (3.13)

and we receive by consideration the equation (B.1) such temporal functions:

$\tau_{c}(\omega,\mathbf{r})=r\ $sgn$\omega$; (3.14)

$\tau^{(1)}(\omega,\mathbf{r})=-ir\ \cot(\omega r)$. (3.15)

The expression (3.14) shows that causal propagators describe free photons
propagations in wave zone and it means that the usual QED calculations
practically exclude phenomena connected with near fields.

The expression (3.15) reveals that response function tends to peaks as $%
\omega r\rightarrow2\pi n$ from below, i.e. at $\tau_{2}^{(1)}(\omega,%
\mathbf{r})\rightarrow-\infty$. Subsequent terms of the expansion (3.1), as
they are expressed through derivatives of $\tau^{(1)}$, only sharpen these
peaks and therefore it can be concluded that the results of QED completely
confirm our main condition (2.18).

\qquad

{\large 3-3. TUNNELING}

Tunnel transitions, as is known, are intimately related with the imaginary
values of momenta. As the demonstrative example of such phenomena the tunnel
transition of particle with energy $E$ through the rectangular barrier $U(x)$
in the 1-D space range $(-a,a)$, where $E<U_{0}$, can be considered.

Energy of particle moving in the potential $U(x)$ is equal to $%
E=p^{2}/2m+U(x)$. For reaching the classically forbidden region, $E<U$,
kinetic energy should be negative, corresponding to imaginary momenta $p$.
In the WKB approximation the wave function $\psi(x)\ \symbol{126}\
\exp(i\Phi(x))$ with $\Phi(x)=\pm\ \int^{x}dx\prime
p(x\prime)/\hbar+O(\hbar) $, where $p(x)=[2m\ (U(x)-E)]^{%
%TCIMACRO{\UNICODE{0xbd}}%
%BeginExpansion
{\frac12}%
%EndExpansion
}$ is the local momentum. In the classical region, the wave function is
oscillatory one, while in the classically forbidden region (corresponding to
imaginary momenta) the wave function is exponentially suppressed.

The matrix element of this transition is of order

$M\ \symbol{126}\ \exp(-\dint _{-a}^{a}[2m\ (U(x)-E)/\hbar^{2}]^{1/2}dx$,
(3.16)

where $a$ is determined from the equality $U(\pm a)=E$. This representation
leads to the following expression for time durations:

$\tau(E)=\hbar\frac{\partial}{i\partial E}\ln M=-$Im$\dint _{-a}^{a}[2m\
(U(x)-E)/\hbar^{2}]^{-1/2}dx$, (3.17)

which can be rewritten via imaginary moment of tunneling particle. Here can
be noted that at the considering of tunnel processes by the Landau method
the transition to an ''imaginary'' time is often employed, but as the formal
procedure only [29].

For a sufficiently common case of the oscillator (parabolic) barrier $%
U(x)=U_{0}-%
%TCIMACRO{\UNICODE{0xbd}}%
%BeginExpansion
{\frac12}%
%EndExpansion
\kappa\ x^{2}$ the expression (3.17) leads to such results:

$\tau_{1}(E)=0$;$\qquad\tau_{2}(E)=-\frac{\pi}{2\omega_{0}}=-\frac{\lambda
_{0}}{4c}\qquad$(3.18)

with an oscillator frequency $\omega_{0}=(k/m)^{1/2}$, i.e. behind the
barrier the complete energy of particle (wave) must appear instantaneously
and regardless of its value, as $\tau_{2}$ in (3.18) corresponds to $%
\lambda/4c$. This property is retained even under energy absorption at the
tunneling state, e.g. at substitution $U_{0}\rightarrow
U_{0}+U_{1}\cos\omega t$, if $U_{0}+U_{1}>E$. Note that $|\omega_{0}%
\tau_{2}| $ is bigger than uncertainty magnitude and hence can be measurable.

Independence from the energy does not represent the common property of
tunnel transitions. So for the rectangular barrier $\tau_{2}(E)=-[2m\
a^{2}/(U-E)]^{1/2}$, and here the advancing duration (extent of instant
jump) varies with $U(t)$.

The expression (3.17) really shows that the tunnel transition proceed
without delays as $\tau_{1}=0$, but $\tau_{2}$ is negative and in general
qualitatively corresponds to the uncertainty value. For $E>U_{0}$,
evidently, $\tau_{1}\neq0$ and both its signs are, in principle, possible.

Let's consider, as a more specific example, the variation of wave function
in this process by direct calculations. Let the process begins with the
Gaussian wave packet, centered around $\mathbf{k}_{1}$, on the left side of
barrier

$\psi_{i}(t,x)=\dint _{-\infty}^{\infty}\ dk\ f(k-k_{1})\exp[i(kx-E(k)t)]$,
(3.19)

The usual calculation of function on the right side of barrier leads to

$\psi_{f}(t,x)\cong\psi_{i}(t,x-2a)$, (3.21)

which shows that the transmission through barrier is instantaneous
(unessential phase factors are omitted).

The strange, contradictory and curious, as seems, effect of instantaneous
packet transfer is usually explained as erroneous, explicable by physical
noncompleteness of nonrelativistic Schr\"{o}dinger equation (e.g. [14]). But
as was shown by different methods, just similar instantaneous jumps are
characteristic for completely relativistic QED expressions. Hence may be
concluded that the possibilities of superluminal or even instant transitions
are the characteristic peculiarity of quantum tunneling.

\qquad

{\large 3-4. FRUSTRATED TOTAL INTERNAL REFLECTION (FTIR)}

For analysis of experiments [7] the basic peculiarities of phenomenon of
FTIR must be considered. As FTIR can be described classically, we can begin
the consideration with the Fresnel formulas. In accordance with them
amplitudes of evanescent waves, tunneling into optically lesser dense medium
under the angle $\vartheta\geq\sin^{-1}(1/n)$, and amplitudes of appropriate
usual waves in $(x,z)$ plane are connected on the borderline $z=0$ by the
response function [30]:

$S_{e}(\omega|t,x,z)=A(n,\vartheta)\exp[-i\omega(t-x\ n\ \sin\vartheta
)-\omega\ s$\/ $z]$, (3.22)

where $s=(n^{2}\sin^{2}\vartheta-1)^{1/2}$, $n>1$, factor $A(n,\vartheta)$
depends on polarization, etc.

The condition of completeness of (3.22) should be expressed as

$\dint _{0}^{\infty}|S_{e}|^{2}dz=1$, (3.23)

which shows that in the description of FTIR must be included all
trajectories of evanescent ''rays'' with $z\in \lbrack 0,\infty )$. But if
durations of their passage depend on $z$, the short initial wave pulse
should be effectively extended after each act of FTIR. However, already the
waveguides practice shows absence of appreciable expansion of short pulses
[31] (just absence of an expansion is the basic one for single mode light
guides). Therefore it should be assumed, that the durations of
''geometrically differing trajectories of rays'' do not depend on $z$ and
these durations should be (approximately, at least) equal, i.e. the
transverse shifts of ''trajectories'' of evanescent rays must be instant:
the duration of signal propagation in $x$ direction in the area of FTIR
should been determined by the speed component $v_{x}=(c/n)\sin \vartheta $
only.

The direct analogy to the consideration carried out above permits to rewrite
(3.22) as $S_{e}(\omega;\mathbf{r})=|S_{e}|\exp(i\Phi)$ with the real phase $%
\Phi(\omega,\mathbf{r})$ and then, according to (3.2),

$\tau_{1}=\partial\Phi/\partial\omega=-(t-n\ x\sin\vartheta)$, (3.24)

$\tau_{2}=\frac{\partial}{\partial\omega}\ln|S_{e}|=-z[s+\frac{\omega}{s}\
n\ \sin^{2}\vartheta\ \frac{\partial n}{\partial\omega}]$. (3.24')

Hence the response function for transparent linear medium and weak light
flux can be written down at $|\omega|\ \symbol{126}\ \omega_{0}$ as

$S(\omega,\mathbf{r})=S(\omega_{0},\mathbf{r})\ \exp\{(\omega-\omega
_{0})[i\ \tau_{1}(\omega_{0},\mathbf{r})-\tau_{2}(\omega_{0},\mathbf{r})]\}$%
. (3.25)

On the other hand, we can begin with (3.25) as with the independent
determination of response function. Then according to [32] the changes of
moment of particle during quantum transition is determined as $\Delta 
\mathbf{p}=\rho\ \left\langle \mathbf{E}\right\rangle \ V\ T_{p}$, where $%
\rho$ is the average density of charges in volume of interaction $V$, $%
\left\langle \mathbf{E}\right\rangle $ is the average intensity of an
internal field, $T_{p}$ is the duration of process. At FTIR the magnitude
and direction of additional moment, received by photon in the area of FTIR,
should be determined by parameters of field $\mathbf{E}$ in immediate
proximity to media interface layer. And actually this formula conducts to
the condition, described above: the performance of (3.22) corresponds to the
transformation of moment, $kz\rightarrow ikz$, and (3.25) corresponds to the
representation of complex temporal function $T_{p}\rightarrow\tau_{1}+i%
\tau_{2}$, which at $\tau_{1}=0$ leads to concordance of both approaches.

Note that on the same basis some other phenomena, analogs of FTIR, can be
considered (their description is given e.g. in [33]).

\qquad

{\Large 4. COMPARISON WITH EXPERIMENTS}

The cited experiments, except FTIR, can be divided onto two groups: 1). The
resonant ones, in which ''gain time'' can be, in principle, measurable after
single scattering act, and 2). Nonresonant (relative to atomic frequencies)
processes of light flux passage through media, where tiny ''gain'' durations
can be accumulated via sequential interactions.

1). The best demonstration of existence of superluminal propagation is
presented in the articles [2]: the weak almost resonant laser pulse of $%
\lambda=0.8521$ mcm was passed through tube of $6$ cm length with atomic
Cesium at temperature of $30^{0}C$. At the deviation of frequency from
resonance into the abnormal dispersion side on $\Delta\nu=1.9$ MHz the light
pulse, propagated through tube, outstrips light pulse, transferred through
vacuum, on $(62\pm1)$ nsec, which can be named ''the gain time''.

The estimation of free path length and almost complete similarity of the
forms of input and output pulses show, that the pulse has undergone only one
act of outstripping scattering in tube. In support of this assumption can be
noted that the output signal is of order $40\%$ of an entrance signal that
corresponds to division of an entrance pulse in each scattering act on
approximately equal retarding and advancing parts. As the width of top level 
$\Gamma=0.37\cdot10^{8}$ $\sec^{-1}$ [20], the estimation of advancing by
the formula (3.12) conducts to the ''gain duration'' $\tau_{eff}\ \symbol{126%
}\ \tau_{2}-1/\omega\ $\symbol{126}$\ -59.4$ nsec in the consent with
measured value of advancing or outstripping.

The experiments [3] are of special interest to suggested theory. In them the
superluminal features of pulses propagated through microcavity containing a
few cold atoms of $^{85}$Rb were observed. Number of atoms was varied and,
in principle, the results could be extrapolated till single atom in the
resonator. The observations of superluminal propagation in such
''substance'' evidently prove that this phenomenon is not caused by
rearrangement of wave front [11-13] and so on, but it must be related to
processes of scattering on individual atoms.

The numerical treatment of these experiments show that with detuning of
initial laser frequency on $53$ MHz the advancing of signal on $170$ nsec is
observed, but with detuning on $45$ MHz it gave way to delay on the $440$
nsec. It means that the strict resonance is above the frequency of
superluminal signal not more than on $8$ MHz. On the other hand the observed
outstrip on $170$ nsec corresponds, in accordance with (2.12), to difference
of frequencies of order of $5.88$ MHz from resonance into the region of
abnormal dispersion.

Hence we should conclude the conformity of these experimental data with
suggested theory.

Now let consider the first, as far as we know, observation of the
superluminal phenomenon [1]. In this experiment it was established that the
pulse of He-Ne laser ($\lambda =0.6328\ $mcm) passes through a tube with $%
^{20}$Ne ($p=2.6$ Torr, $L=16$ cm) with the group velocity $u=1.0003c$.

Let's try to be limited by consideration of this experiment to account of
the closest resonance on $\lambda_{21}=0.6334$ mcm with $A_{21}=1.36%
\cdot10^{7}$ sec$^{-1}$ only. Density of atoms in the pipe with such
pressure of gas is of order $N_{a}=9.2\cdot10^{16}$ atoms/cm$^{3}$ (i.e. $%
N_{e}\ \symbol{126}\ 8N_{a}$ electrons of upper atomic shell considered as
scatterers). The cross-section of unbiased elastic scattering $\sigma=16\pi\ %
\left[ \frac{2j_{2}+1}{2(2j_{1}+1)}\right] \ \delta\omega\ (\frac{c}{\omega}%
)^{2}|\tau_{eff}|\approx3\cdot10^{-20}$ cm$^{2}$. Hence the free path length 
$l=1/N_{e}\sigma=45$ cm and each photon cannot suffer more than one
scattering act. Thereby the ''gain duration'' should be calculated for
single atoms: as the difference of frequencies is of order of $453$ GHz, it
leads, in accordance with (3.12), to the advancing on $0.35$ psec. These
values gives $u/c\ \symbol{126}\ 1.0006$, qualitatively corresponding to the
results of [1].

2). Let's consider the experiments, in which the superluminal phenomena in
nonresonant (relative to atomic levels) conditions were observed. For such
cases all photons can be divided on ''superluminal'' and usual ones. For
''superluminal'' photons, which have experienced consecutive outstripping
(advanced) interactions only, the average number of such interactions on
distance $L$ will be of order $N=L/(l+\Delta l)$, where $l=1/\rho\sigma$ is
the free path length, $\rho$ is the density of scatterers, $\sigma$ is the
complete cross-section of $e$-$\gamma$ scattering and $\Delta l=c|\tau_{2}|\ 
$for $\tau_{2}<0$. Therefore, the durations of light flux flight through
vacuum and superluminal photons flight and jumps through medium are equal,
accordingly, to

$T=L/c$;$\qquad T_{adv}=(L-N\Delta l)/c\rightarrow T-\Delta T$. (4.1)

It results in an obvious expression for average speed of ''superluminal''
photons:

$u/c=T/T_{adv}=1/(1-c\rho\sigma|\tau_{2}|)$. (4.2)

In the experiments [5] superluminal propagation of pulse of intensity $3\div
100$ W/cm$^{2}$ through the film of GaP:N with changes of thickness of film
in an interval $9.5\div 76$ mcm was investigated with variation of the laser
radiation frequency around the isolated $A$-line ($534$ nm). The received
diagram of dependence of duration of delay (positive and negative) as
function of frequency of light qualitatively corresponds to the expression
(3.4) with the dramatic transition from subluminal to superluminal speeds.

In the series of experiments [6] passage of light pulse through multilayer
dielectric mirrors was investigated. The mirrors contained alternating
layers of thickness $\lambda/4$ with high ($H$) and low ($L$) indices of
refraction, so they were of structure $(HL)^{m}H$, in whole were
investigated mirrors with $m=3\div11$.

From the classical point of view each pair of layers should completely
reflect falling resonant wave. But as some photons are passed, due tiny
superluminal jumps, for those times on larger distances, they slip through
interferential reflecting planes. Here can be \textit{accepted} that as at
nonresonant scattering two Feynman graphs (retarded and advanced) lead to
approximately equal contributions, so each pair of layers approximately
halves passed photon flux into reflected and transmitted parts. Therefore,
if the photon free path length $l\ <\lambda/4$, intensity of light, missed
through such mirror, should contain an outstripping flow of intensity $%
J_{adv}(m)\ \symbol{126}\ J_{0}/2^{m}$ that corresponds to measurements.

As the examined process is nonresonant, then according to principle of
uncertainty or (3.4), where far from all self frequencies the second term
can be omitted, we accept that $\tau_{2}=-1/2\omega$, and it means, that the
process of scattering occurs on almost free electrons. In the same
approximation, according to the optical theorem of scattering theory,
complete cross-section of interaction $\sigma_{tot}=(4\pi c/\omega)\ r_{0}$, 
$r_{0}=e^{2}/mc^{2}$. Thus at density of external electrons $\rho
=1.3\cdot10^{21}$ cm$^{-3}$, the relation (4.2) conducts to $u/c=1.56$, that
corresponds to the experimental data.

The experiments in the microwave range [4] are more difficult for analyzing,
as in them the role of boundary conditions can be essential. Thereby we
shall limit ourselves by qualitative consideration of superluminal
propagation noticed at pass of GHz waves by air between two aerials. If such
process can be described by instantaneous jump of wave on $\Delta l$\
(outstripping radiation) at distance $x$ between megaphones or aerials, it
becomes necessary to expect change of relative speed as $u/c=1/[1-\Delta
l/x] $ and analogical dependence was observed in these experiments.

It is necessary to notice, that the independence of all these effects on
polarization is caused, obviously, by absence of conductors on the way of
light fluxes. The interesting qualitative results are submitted in [8],
where was investigated the diffraction of THz waves on thin wires and plates
and was received, that the superluminal phenomena are appreciable at
polarization parallel to conductors and are absent or are not appreciable at
perpendicular polarization. These results seem explainable, as waves of
parallel polarization should generate an alternating current in a long
conductor, which thereby becomes the radiating aerial with the extended near
field zone. In case of perpendicular polarization the induction is much
weaker and diffracted beams of far field are imposed on weak radiation of
near field.

Let discuss the experiments on FTIR.

In the experiments [7 a)] ''gain duration'' (3.24') in the area of FTIR was
measured in THz region: the distribution of wave packages with central wave
length of $1$ mm and pulse duration of $0.8$ psec was investigated at depths
of FTIR from $0$ up to $8$ mm. The researches showed outstripping character
of evanescent waves, the numerical estimation of this advancing, obviously,
linearly depends on the size of an interval, in which these waves are
observed. According to (3.24') and if omit the term $\partial $n/$\partial
\omega $, the observed data conducted to $\tau _{2}=-0.41$ psec on $1$ mm of
depth of evanescent waves penetration.

In the experiments [7 b)] dependences of outstripping of evanescent waves on
depth $d$ was measured at various light polarizations ($\lambda=3.39$ mcm, $%
d=0\div25$ mcm) and it has been shown, that the duration of an advancing
does not depend on polarization. Up to distance $d=8\ $mcm it grows and then
remains approximately constant, of order of $0.2$ psec. These supervisions
do not contradict our approach, but do not give opportunities of
quantitative comparison.

\qquad\qquad

{\Large 5. DISCUSSION}

As the opportunity of instantaneous propagation of signal (or excitation)
within a near field zone, its transferring, is established and as it is
demonstrated that by this phenomenon can be explained a set of experimental
facts, connected with ''gain times'', then now the attempts to consider the
problem with more general positions become possible.

Let's examine some theoretical prerequisites and possibilities for
interpretation of the deduced results.

1. It is necessary to emphasize, at the beginning, that the principle of
locality experimentally was checked up only in the far field, only for
fields $\mathbf{E}$ and $\mathbf{H}$. Hence, the \textit{a priori} excluding
of possibilities of nonlocality of those parts of electromagnetic field,
which are not included into the (transverse) far electrical and magnetic
fields, represents not obvious hypothesis.

The hidden opportunity of nonlocality ''in small'' can be contained in
conditions of gauge invariance: the classical Lorentz condition, $\partial
A_{\mu}/\partial x_{\mu}=0$, is replaced in QED by the Lorentz - Fermi
condition [34]:

$\frac{\partial A_{\mu}}{\partial x_{\mu}}\left| 0\right\rangle =0$, (5.1)

which requires a mutual indemnification or disappearance of the
''superfluous'' components of $A_{\mu}$, ''pseudophotons'', \textit{only in
the average}. Therefore it does not exclude possible nonlocality of
interaction in a near zone of these fields.

All attempts of fields' quantization without introduction of 4-vector $%
A_{\mu }$ were unsuccessful. Then the Aharonov-Bohm effect brightly showed,
that these ''needless'' field components have definite physical sense and
the impossibility of their omitting has, apparently, deep roots. This
conclusion could be strengthened by existence and features of Casimir forces
and, probably, by the phenomena of near field optics (e.g. the review [35]).

It can be added also, that consideration of some concrete phenomena
connected to longitudinal part of electromagnetic field (cf. [36]) leads to
phenomenological introduction of nonlocality, of some effective ''smearing''
of charge, depending on its 4-moment.

So, the problem of locality of these components of field is not resolved yet.

2. The consideration of all these phenomena invites further investigations
of the basic equations of theory. So, if to examine the Klein-Gordon equation

$[\partial_{t}^{2}-\Delta-m^{2}]\ f(x)=0$, (5.2)

as the wave equation, i.e. to search its solution as

$f(x)=f(vt-r)$, (5.3)

then non-fading decision exists only at $v\leq c$. Just this circumstance
forbids movement with superluminal speed as effective nonlocality on any
distances. However there is not formal restriction on the existence of
non-wave solutions in a scope of near field with arbitrary value of $v$. So
it is necessary to consider such solutions, which can transfer interaction
via field $A_{\mu}$ with speed greater $c$ on small distances, i.e.
effectively nonlocal.

3. Let's point out such formal opportunity of interpretation of the
equations describing processes faster-than-$c$. Projectors in (3.6-7) can be
expressed through the Green functions for Klein-Gordon equation with
imaginary mass: for example,

$P_{NE}(\omega,\mathbf{k})=\theta(-k^{2})=\dint _{0}^{\infty}dm^{2}\
\delta(-\omega^{2}+\mathbf{k}^{2}-m^{2})=2\dint
_{0}^{\infty}dm^{2}\Delta_{1}(\omega,\mathbf{k},im)$, (5.4)

i.e. it formally describes interaction transferred by tachyons, hypothetical
particles with velocity always-bigger $c$. From here for (3.7) in the
complete x-representation follows such dispersion relation:

$f_{NE}(x)=2\pi\dint _{0}^{\infty}dm^{2}\dint d^{4}y\ \Delta_{1}(x-y,im)\
f_{NE}(y)$. (5.5)

For functions $f_{E}(x)$ is received the same representation, but with real
mass.

Thus, it is not excluded a formal possibility for description the
superluminal phenomena via tachyons (cf. [37]).

4. At transition to imaginary time, e.g. at transformation of variables $(t,%
\mathbf{r})\rightarrow(i\tau_{2},\mathbf{r})$ at $\tau_{1}=0$, the Equation
(5.2) transforms from hyperbolic into elliptic one, in the 4-D Poisson
equation:

$[\partial_{\tau}^{2}+\nabla^{2}]$ $D_{c}(i\tau,\mathbf{r})=-\delta(x)$,
(5.6)

and hence, all signals in such field are transferred instantly: the field
becomes instanton ones (cf. [38]). It means that the photon, absorbed on one
side of such pseudoparticle of radius $\Delta l=c|\tau_{2}|$, instantly
appears on the opposite side. Thus instead discussion of the advanced
character of emission, instantaneous jumps and so on, we can take into
account the existence of virtual instanton-type states (of zero mass in
considered case) with parameters specified by the usual propagators via time
duration terms.

Such description seems sufficiently simple and physically reasonable.

But a somewhat more complicated will be the structure of near field in the
FTIR phenomenon. Its description by the substitution $kz\rightarrow ikz$
into the Fresnel formulae leads to replacement of the usual ansatz $%
Q_{n}(k)=k_{0}^{2}-n^{2}\mathbf{k}^{2}$ for the far field on ansatz $%
Q_{e}(k)=k_{0}^{2}+n^{2}(k_{z}^{2}-k_{x}^{2}-k_{y}^{2})$ for the FTIR zone.
It means transition from the wave (hyperbolic) equation into the
ultrahyperbolic equation [39] for near-surface evanescent waves:

$L(ix)\
S_{e}\equiv(\partial_{t}^{2}+\partial_{z}^{2}-\partial_{x}^{2}-%
\partial_{y}^{2})\ S_{e}=\pm\ \delta(x)$. (5.7)

This equation can be considered as the difference of two 2-D Laplace
equations $(\Delta_{t,z}-\Delta_{x,y})\ f(x)=\delta(x)$ with Green functions
(propagators)

$G^{\pm}(x)=1/(L(x)\pm\ i0)\equiv1/(t^{2}+z^{2}-x^{2}-y^{2}\pm\ i0)$. (5.8)

Their difference should been the Green function of corresponding homogeneous
equation. However, as the full Fourier-transforms \textbf{F}$_{k}[G^{\pm
}(x)]=4\pi/(-G(k)\pm\ i0)$ of both functions (5.8) coincide, the homogeneous
form of equation (5.7) has only the zero general solution. It means the
absence of free waves, which would propagate in $z$-direction. On the other
hand, it means that the $x$-component of light flux speed at FTIR depends on 
$k_{x}$ component of moment only, and for this reason, as was noted above,
light pulses at FTIR are not widening.

However, for complete reliance in their existence the further researches are
necessary.

\qquad

{\Large CONCLUSIONS}

The received results demonstrate, in accordance with the experimental data,
the reality of superluminal signal transferring. Hence they prove the
validity of Wigner's general principle of causality with its admitting of
instantaneous transferring. It seems that these reasonings can be considered
as the first proof of existence of instantons as virtual particles or, more
precisely, as pseudoparticles.

The principle of causality is refined, and it is shown that ''sizes of
scatterers'' or ''sizes of scattering processes'', often introduced in
different investigations, can be identified, at least in definite cases,
with the instanton sizes. It is not excluded that just such instantaneous
transferring can take place at transfer of excitations or even binding
energy between some constituents of condensed media (the F\"{o}rster law and
so on). Hence some of their parameters can be defined by the extent of area
of tunneling (probably, in the scope of near field).

All it shows the necessity of (formal) refinement of the first postulate of
relativity: \textit{the speed of signal propagation in far field cannot
exceed the vacuum velocity of light, but the transferability of excitation
can be instantaneous on the length of tunneling (within the scope of near
field)}. From the kinematical point of view such transitions can be
described via magnitudes of ''gain time'' at processes of particle (wave)
propagation. Note that the physical sense, as was discussed above, reveals
that the expression ''superluminal transferring'' is more exact and
preferable than the usual now ''superluminal propagation'' and so on.

It can be pointed also, that the physical sense of instantaneous character
of transition consists, in particular, in the answer on very old naive
question: where resides particle at the time of tunnel transition? And also:
how can be imaged a process of gradually exit of emitting wave (on its
length, at least) from the source?

On the other hand, it must be underlined that in the chain of these
investigations had been demonstrated the significance and possibilities of
theory of temporal functions, at least in their specific forms. Moreover, it
must be noted the suggesting and developing of the specific mathematical
method of equations of orthogonality, which will be more comprehensively
considered elsewhere.

\qquad

{\large ACKNOWLEDGMENTS}

The author wishes to thank I.I.Royzen for many critical comments,
M.Ya.Amusia, R.Englman, R.E.Kris, G.Nimtz and G.M.Rubinstein for valuable
discussions and support.

\qquad

{\Large APPENDIX}

\textbf{A}.\textbf{\ CONDITIONS OF COMPLETENESS AND EQUATIONS OF
ORTHOGONALITY }

The most serious assumption at deduction of conditions of orthogonality
(2.3-4) and others consists in essentially used classical condition of
completeness:

$P_{L}+P_{NL}=1$, (A.1)

Really it means an implicit introduction of the assumption about absence of
features connected to intersection of areas, $P_{L}\bigcap P_{NL}=0$, and
requires special researches. Notice that the Heaviside unit functions $%
\theta(\xi)$, from which projectors are constructing, are not definite in
the point $\xi=0$; the considered problem, just as many others in the
theory, is intimately connected with this uncertainty.

The decomposition (A.1) can be generalized as

$P_{L}+P_{NL}+P_{\delta L}=1$, (A.2)

into which are introduced the ''quasilocal terms'' (cf. [40]):

$P_{\delta L}=a_{0}\delta(x^{2})+a_{1}\partial_{\mu}^{2}\delta(x^{2})+...\ $%
. (A.3)

At limiting by the first term of (A.3) and as in the $(t,\mathbf{k})$%
-representation the Fourier transform $\mathbf{F}_{\mathbf{k}}[\delta
(x^{2})]=\frac{1}{2\pi^{2}k}\sin(k|t|)$, we receive that the argument of $%
\delta$-function in (2.11) is replaced on

$\frac{\cos(kt)}{2\pi^{2}k^{3}}[(1+a_{0}k^{2})\tan(kt)-kt]$ (A.4)

and consequently the transcendental equation, which roots determine
properties of nonlocal functions, is slightly varied.

However, insofar our main result, the conclusions about instantaneous
excitation transferring by tunneling, does not change, we can do no more
than mention a possibility of consideration of quasilocal terms in
decomposition of projectors and accordingly in decomposition of response
functions.

\qquad

\subparagraph{\textbf{B. TO INTRODUCTION AND DETERMINATION OF TEMPORAL
FUNCTIONS}\qquad\ }

The designation (3.2) can be rewritten as the equation for $S$-matrix:

$(\partial/i\partial\omega)\ S(\omega,\mathbf{r})=\tau(\omega,\mathbf{r})\
S(\omega,\mathbf{r})$. (B.1)

Logically another way seems more reasonable: the transition to such equation
by the complete Legendre transformation $(t,\mathbf{r})\rightarrow (\omega,%
\mathbf{k})$ of the Schr\"{o}dinger equation for $S$-matrix:

$(i\partial/\partial t)\ S(t,\mathbf{r})=\mathbf{H}(t,\mathbf{r})\ S(t,%
\mathbf{r})$, (B.2)

This way leads, instead of (B.1), to somewhat different form:

$(\partial/i\partial\omega)\ S(\omega,\mathbf{k})=\mathbf{T}(\omega ,\mathbf{%
k})\ S(\omega,\mathbf{k})$ (B.3)

with temporal function instead of Hamiltonian.

It should be noticed that just these functions arise into theory in the
quite reasonable way: they are naturally revealed in the standard QED
calculations [41]. That is the reason that we did not discuss here many
other suggested forms of time-delays, e.g. [42].

Fourier transformation of (B.1),

$(\partial/i\partial\omega)\ S(\omega,\mathbf{k})=\int d\mathbf{q}\tau
(\omega,\mathbf{q})\ S(\omega,\mathbf{k-q})$, (B.4)

shows that both definitions are rather close if it can be suggested that the
main role in discussed processes play closely related magnitudes of wave
numbers. (These items will be discussed more comprehensively elsewhere.)

It must be noted that the first approaches and views on the problem require,
in accordance with a common intuition, consideration of their relation to
the uncertainties principles. Therefore must be considered peculiarities of
uncertainty principle, connected with temporal functions [43].

Let's use the general method of deduction of uncertainties relations given
by Schr\"{o}dinger [44]. It starts with decomposition of the product of
operators on Hermitian and anti-Hermitian parts as $\mathbf{AB=%
%TCIMACRO{\UNICODE{0xbd}}%
%BeginExpansion
{\frac12}%
%EndExpansion
(AB+BA)+%
%TCIMACRO{\UNICODE{0xbd}}%
%BeginExpansion
{\frac12}%
%EndExpansion
(AB-BA)}$ with subsequent quadrates of this expression, its averaging over
the complete system of $\psi$-functions and replacement of operators by
differences of operators and their averaged values: $\mathbf{A\rightarrow
A-\ }\left\langle \mathbf{A}\right\rangle $. This leads, due the Schwartz
inequality, to such final expression:

$(\Delta A)^{2}(\Delta B)^{2}\geq%
%TCIMACRO{\UNICODE{0xbc}}%
%BeginExpansion
{\frac14}%
%EndExpansion
\mathbf{|}\left\langle \mathbf{AB-BA}\right\rangle \mathbf{|}^{2}\ \mathbf{%
+\ 
%TCIMACRO{\UNICODE{0xbc}}%
%BeginExpansion
{\frac14}%
%EndExpansion
(}\left\langle \mathbf{AB+BA}\right\rangle \ \mathbf{-\ }2\left\langle 
\mathbf{A}\right\rangle \left\langle \mathbf{B}\right\rangle \mathbf{)}^{2}$%
, (B.5)

where to the common form became added the last term, which is an Hermitian
function and its magnitude must be real. Note that the Heisenberg limit of
(B.5) shows a minimal value of uncertainties, which can be achieved at the
determined conditions.

In considered case the canonically conjugated operators must be substituted
by differences $\mathbf{A\rightarrow E-}\left\langle \mathbf{E}\right\rangle 
$ and $\mathbf{B\rightarrow T-}\left\langle \mathbf{T}\right\rangle $ and
there is needed their averaging, instead of $\psi$-functions, over the
complete system of $S(E)$ functions, nonunitary in general, as

$\left\langle \mathbf{A}\right\rangle =\dint _{-\infty}^{\infty}dE\ S^{\ast}%
\mathbf{A}S\ /\dint _{-\infty}^{\infty}dE\ |S|^{2}$. (B.6)

The evident calculations in $E$-representation with operators $\mathbf{T}$
and $\mathbf{E}$ give such result:

$(\Delta E)^{2}(\Delta T)^{2}\geq%
%TCIMACRO{\UNICODE{0xbc}}%
%BeginExpansion
{\frac14}%
%EndExpansion
\ \hbar^{2}+%
%TCIMACRO{\UNICODE{0xbc}}%
%BeginExpansion
{\frac14}%
%EndExpansion
\ (\left\langle \mathbf{E}\tau_{1}\right\rangle -2\left\langle \mathbf{E}%
\right\rangle \left\langle \tau_{1}\right\rangle )^{2}$, (B.7)

i.e. the general form of uncertainty principle does not depend on the
duration of state formation or on the ''gain time'' at jumps, if $\tau _{2}$
is negative. It can means that peculiarity of $\tau _{2}$ must be considered
as the internal property of forming particle, which does not depend on
measurement procedures. \bigskip \qquad 

{\large REFERENCES}\bigskip 

*). E-mails: perelman@vms.huji.ac.il, mark\_perelman@mail.ru\qquad 

[1]. F.R.Faxvog, C.N.Y.Chow, T.Bieber, Y.A.Carruthers. Appl.Phys.Lett., 
\textbf{17}, 192 (1970).

[2]. L.J.Wang, A.Kuzmich, A.Dogariu. Nature, \textbf{406}, 277 (2000);
A.Dogariu, A.Kuzmich, L.J.Wang. Phys.Rev.A, \textbf{63}, 053806 (2001). .

[3]. Y.Shimizu, N.Shiokawa, N.Yamamoto, M.Kozuma, T.Kuga, L.Deng,
E.W.Harvey. Phys.Rev.Lett., \textbf{89}, 233001 (2002).

[4]. a). A.Enders, G.Nimtz. J.Phys.I (France), \textbf{2}, 1693 (1992); 
\textbf{3}, 1089; \textbf{4}, 565 (1993); A.Enders, G.Nimtz, H.Spieker.
J.Phys.I (France), \textbf{4}, 1379 (1993); A.Enders, G.Nimtz. Phys.Rev.E, 
\textbf{48}, 632 (1994); Phys.Rev.B, \textbf{47}, 9605 (1994); G.Nimtz. In:
arXiv:physics/02040403 v1 16 April 2002;

b). A.Ranfagni, P.Fabeni, G.P.Pazzi, D.Mugnai. Phys.Rev.E, \textbf{48}, 1453
(1993); A.Ranfagni, D.Mugnai. Phys.Rev.E, \textbf{54}, 5692 (1996);
D.Mugnai, A.Ranfagni, R.Ruggeri. Phys.Rev.Lett., \textbf{84}, 4830 (2000).

For comparison of these groups of experiments see: F.Cardone, R.Mignani.
Phys.Lett. A, \textbf{306}, 265 (2003).

[5]. S.Chu, S.Wong. Phys.Rev.Lett., \textbf{48}, 738 (1982).

[6]. A.M.Steinberg, P.G.Kwiat, R.Y.Chiao. Phys.Rev.Lett., \textbf{71}, 708
(1993); Ch.Spielmann, R.Szipocs, A.Stingl, F.Krausz. Phys.Rev.Lett., \textbf{%
73}, 2308 (1994); A.M.Steinberg, R.Y.Chiao. Phys.Rev.A, \textbf{51}, 3525
(1995).

[7]. a). Ph.Balcou, L.Dutriaux. Phys.Rev.Lett., \textbf{78}, 851 (1997); b).
J.J.Carey, J.Zawadzka, D.A.Jaroszynski, K.Wynne. Phys.Rev.Lett., \textbf{84}%
, 1431 (2000).

[8]. K.Wynne, D.A.Jaroszynski. Opt.Lett., \textbf{24}, 25 (1999).

[9]. I.Alexeev, K.Y.Kim, H.M.Milchberg. Phys.Rev.Lett., \textbf{88}, 073901
(2002).

[10]. M.W.Mitchell, R.Y.Chiao. Am.J.Phys., \textbf{66}, 14 (1998);
T.Nakanishi, K.Sugiyama, M.Kitano. Am.J.Phys., \textbf{70}, 1117 (2002). See
also: R. Y. Chiao, J. M. Hickmann, C. Ropers, D.Solli. In: N. Bigelow
(Org.). \textit{Coherence and Quantum Optics} VIII. 2002.

[11]. G.Nimtz, W.Heitman. Progr.Quant.Electr., \textbf{21}, 81 (1997);
R.Y.Chiao, A.M.Steinberg. Phys.Scr. T, \textbf{76}, 61 (1998); E.Recami.
Found.Phys., \textbf{3}1, 1119 (2001); A.M.Steinberg. In: \textit{Time in
Quantum Mechanics }(J.G.Muga e.a., Ed's). Springer, 2002, p.p.305-325;
P.W.Milonni. J.Phys.B., \textbf{35}, R31 (2002).

[12]. P.Mittelstaedt, G.Nimtz (Ed's). Workshop on Superluminal Velocities,
Cologne, 1998. Ann.Phys. (Leipzig), \textbf{7}, 585-782 (1998); R.Bonificio
(Ed.). \textit{Mysteries, Puzzles and Paradoxes in Quantum Mechanics}. (AIP
Conf. Proc. 461). AIP: NY, 1999.

[13]. In the most part of these and other analogical theoretical models are
examined problems of rearrangement of wave fronts by interference effects
and possibilities of comparison of those models with relativistic causality.
Such ideas are not compatible with the observations [3], and as we will
arrive at another conclusions, we cannot compare our results with these
models. On the other hand their complete critical examination is far from
our purposes, the possibilities of tachyon nature of these effects are
discussed below.

[14]. L.A.MacColl. Phys.Rev., \textbf{40}, 621 (1932). More usual
presentation: E.O.Kane. In: \textit{Tunneling Phenomena in Solids}.
(E.Burstein, S.Lundqvist, Ed's). NY: Plenum, 1969. Further references, e.g.:
M.Morgenstern et al. Phys.Rev.B, \textbf{63}, 201301 (2001), P.Krekora,
Q.Su, R.Grobe. Phys.Rev.A, \textbf{63}, 032107; \textbf{64}, 022105 (2001).

[15]. E.P.Wigner. Phys.Rev., \textbf{98}, 145 (1955)

[16]. M.E.Perel'man. Bull. Israel Phys.Soc., \textbf{46}, 133 (2000).

[17]. Y.Aharonov, D.Bohm. Phys.Rev., \textbf{115}, 485 (1959); the review:
S.Olariu, I.Popescu. Rev.Mod.Phys., \textbf{57}, 339 (1987).

[18]. Reviews in the special issue of Comm. At. Mol. Phys., D2, 171-382,
2000.

[19]. J. von Neumann. \textit{The Mathematical Foundation of Quantum
Mechanics}, (Dover, New York, 1959)

[20]. A.Messiah. \textit{Quantum Mechanics}. Vol. 1. Interscience, 1961;
A.Bohm. \textit{Quantum Mechanics: Foundations and Applications}. Springer,
1986.

[21]. M.E.Perel'man. Bull. Acad. Sc. Georgian SSR, \textbf{62,} 33 (1971)
[Math. Rev., \textbf{47}, 1776 (1974)].

[22]. M.E.Perel'man. Zh.Eksp.Teor.Fiz., \textbf{50}, 613 (1966) [Sov. Phys.
JETP, \textbf{23}, 487(1969)]; Dokl. Ak. Nauk SSSR, \textbf{187}, 781 (1969)
[Sov. Phys. Doklady, \textbf{14},772 (1969)]; Bull. Acad. Sc. Georgian SSR, 
\textbf{81}, 325 (1976).

[23]. M.E.Perel'man, R.Englman. Mod. Phys.Lett. B, \textbf{14}, 907 (2000).

[24]. M.E.Perel'man, I.I.Royzen. Bull. Israel Phys.Soc., \textbf{48}, 21
(2002).

[25]. F.T.Smith. Phys.Rev., \textbf{118}, 349 (1960). The Wigner-Smith
determination of delay duration is the most comprehensively elaborated and
presented in: M.L.Goldberger, K.M.Watson. \textit{Collision Theory}. Wiley:
NY, 1964. The recent review: C.A.A.de Carvalho, H.M.Nussenzweig. Phys.Rep., 
\textbf{364}, 83 (2002).

[26]. E.Pollak, W.H.Miller. Phys.Rev.Lett., \textbf{53}, 115 (1984);
E.Pollak. J.Chem.Phys., \textbf{83}, 1111 (1985).

[27]. L.D.Landau, E.M.Lifshitz. \textit{Electrodynamics of continuous media}%
. L.: Pergamon (any edition).

[28]. A.Wheeler, R.P.Feynman. Rev.Mod.Phys., \textbf{17}, 157 (1945), 
\textbf{21}, 425 (1949).

[29]. L.D.Landau, E.M.Lifshitz. \textit{Quantum mechanics}. 2$^{nd}$ ed. NY:
Pergamon, 1965.

[30]. M. Born, E.Wolf. \textit{Principles of Optics}. Cambridge (any
edition).

[31]. J.L.Agudin. Phys.Rev., \textbf{171}, 1385 (1968); G.M.Rubinstein,
M.E.Perel'man. Quant.Electr., \textbf{1}, 983 (1974).

[32]. N.Bohr, L.Rosenfeld. Kgl.Danske Videnk Selskab., Math.-Fys.Medd., 
\textbf{12}, No 8, (1933).

[33]. W.Nasalski. J.Opt.Soc.Am.A, \textbf{13}, 172 (1996) and references
therein

[34]. E.Fermi. Rev.Mod.Phys., \textbf{4}, 87 (1932).

[35]. B.Hecht e.a. J.Chem.Phys., \textbf{112}, 7761 (2000) and references
therein.

[36]. S.Ismail-Beigi et al. Phys.Rev.Lett, \textbf{87}, 087402 (2001) and
references therein.

[37]. S.Liberati, S.Sonego, M.Visser. Ann.Phys. (NY), \textbf{298}, 167
(2002) and references therein.

[38]. R.Rajaraman. \textit{Solitons and instantons}. North-Holland: 1982 and
references therein.

[39]. I.M.Gel'fand, G.E.Shilov. \textit{Generalized functions}. Vol.1.
Moscow, 1959.

[40]. N.N.Bogoliubov, D.V.Shirkov. \textit{Introduction to the Theory of
Quantized Fields}. 3rd ed. Wiley, 1980.

[41]. M.E.Perel'man. Phys.Lett., \textbf{32}A, 64, 1970; Zh.Eksp.Teor.Fiz., 
\textbf{31}, 1155 (1970) [Sov.Phys.JETP, \textbf{31}, 1155 (1970)]; Dokl.
Ak. Nauk SSSR, \textbf{214}, 539 (1974) [Sov.Phys. Doklady, \textbf{19}, 28,
1974]; \textit{Kinetical Quantum Theory of Optical Dispersion}. Tbilisi:
Mezniereba, 1989, 120 p. (In Russian); In: \textit{Multiphoton Processes}
(G.Mainfray \& P.Agostini, Ed's). Paris, CEA, 1991, 155.

[42]. Reviews and discussions in: \textit{Time in Quantum Mechanics}
(J.G.Muga e.a., Ed's). Springer, 2002.

[43]. The deduction of uncertainty principle with operator $\mathbf{T}%
=\hbar\partial/i\partial E$ was performed by E.P.Wigner: \textit{Aspects of
Quantum Theory}. (B.Salam, E.P.Wigner, Ed's). Cambridge, 1972, p.237.

[44]. E. Schr\"{o}dinger. Sitzugsber.Preuss.Akad.Wiss., 1930, S. 296.

\end{document}